\documentstyle[epsfig]{mn}

\def\simlt{\mathrel{\rlap{\lower 3pt\hbox{$\sim$}}\raise 2.0pt\hbox{$<$}}}
\def\simgt{\mathrel{\rlap{\lower 3pt\hbox{$\sim$}} \raise 2.0pt\hbox{$>$}}}
\def\lsim{\mathrel{\rlap{\lower 3pt\hbox{$\sim$}}\raise 2.0pt\hbox{$<$}}}
\def\gsim{\mathrel{\rlap{\lower 3pt\hbox{$\sim$}} \raise 2.0pt\hbox{$>$}}}

\begin{document}
\title[SGRBs: a bimodal origin?]{Short Gamma Ray Bursts: a bimodal origin?}

\author[Salvaterra et al.]{R. Salvaterra$^1$, A. Cerutti$^1$, G. Chincarini$^{1,2}$, 
M. Colpi$^1$, C. Guidorzi$^{1,2}$, \&  P. Romano$^{1,2}$\\
$1$ Dipertimento di Fisica G.~Occhialini, Universit\`a degli Studi di Milano
Bicocca, Piazza della Scienza 3, I-20126 Milano, Italy\\
$2$ INAF, Osservatorio Astronomico di Brera, via E. Bianchi 46, I-23807 Merate
(LC), Italy}

\maketitle \vspace {7cm}

\begin{abstract}
Short--hard Gamma Ray Bursts (SGRBs) are currently thought to
arise from gravitational wave driven coalescences of double 
neutron star systems forming either in the field or dynamically
in globular clusters. For both channels we fit the peak flux 
distribution of BATSE SGRBs to derive the local burst formation
rate and luminosity function. We then compare the resulting
redshift distribution with {\it Swift} 2-year data, showing
that both formation channels are needed in order to reproduce
the observations. Double neutron stars forming in globular clusters are found
to dominate the distribution at $z\lsim 0.3$, whereas the field
population from primordial binaries can account for the high--$z$
SGRBs. This result is not in contradiction with the observed
host galaxy type of SGRBs.
\end{abstract}


\section{Introduction}

The afterglows of several Short Gamma Ray Bursts (SGRBs)
have recently  been localised on the sky with {\it Swift} (Gehrels et al.
2004) and {\it HETE-II} (Lamb et al. 2004), allowing for the 
determination of
their redshift and host galaxy (Nakar 2007; Berger et al. 2007a, and
references therein). 
For SGRBs, the data give  mean redshift $z\sim 0.6,$ and  
early as well as late type galaxy hosts.

The current idea on the nature of the SGRBs is that they arise
from gravitational wave  driven mergers of neutron star-neutron star 
and/or neutron star-black hole binaries (Belczynski et al. 2006;  Nakar 2007). 
Double neutron star binaries (DNSs) are observed in the field of
the Milky Way and in globular clusters, so  they can form through
two different channels: (i) evolution of primordial massive star
binaries, formed in the galactic field (field scenario; Narayan,
Paczynski \& Piran 1992), and (ii) dynamical formation\footnote{Formation 
of primordial DNSs may not be efficient in globular clusters (Ivanova et al 2008). }, 
through
three-body interactions in globular clusters (GCs) involving the exchange of
a light star companion of a neutron star with an incoming isolated
neutron star (GC scenario; Grindlay, Portegies Zwart \& Mc Millan
2006). 

In this Letter, we derive the local SGRB formation rate and luminosity 
function by fitting the peak flux distribution of the large sample of BATSE 
SGRBs (Paciesas et al. 1999) exploring both formation scenarios. 
We compare the model results with
the redshift distribution of the sample of {\it Swift}
SGRBs in the first two years of operation,
with the aim of disentangling the relative importance of the
two SGRB populations at different cosmic epochs.

\section{SGRB formation models}

In this section we compute the intrinsic formation history of SGRBs 
for the two scenarios: the field scenario, where DNSs 
correlate with the overall star formation and the dynamical one inside GCs.

In the field scenario, the intrinsic formation rate, defined as the
number of bursts per unit time and unit comoving volume at redshift
$z$, $\Psi_{\rm SGRB}(z),$ is given by the convolution of the massive
star binary formation rate and a delay time distribution function
$f_{\rm F}(t)$. The delay is the time interval between the formation
of the massive star binary and the merging of the DNSs due to
gravitational wave  emission.  The formation rate is assumed to be
proportional to the cosmic star formation rate, $\dot{\rho}_\star$,
parametrized as in Hopkins \& Beacom (2006). $\Psi_{\rm
SGRB}(z)$ is given by

\begin{equation}
\Psi_{\rm SGRB}(z) \propto \int_z^{\infty}dz' \frac{dt}{dz'}(z') 
\dot{\rho}_\star(z') f_{\rm F}[t(z)-t(z')],
\end{equation}
where $t(z)$ is the age of the Universe at redshift $z$. 
For the delay time distribution function $f_{\rm F}(t)$, we adopt the
following simple expression: $f_{\rm F}(t) \propto 1/t$, as suggested 
from an updated analysis of double compact object mergers
performed using population synthesis methods (Portegies Zwart \& Yungeson 
1998, Schneider et al. 2001, Belczynski et al. 2006, O'Shaughnessy et al. 2008).
Moreover this form seems to be supported
by the distribution of the orbital parameters of six DNSs observed
in our Galaxy (Champion et al. 2004).
Characteristic delay times vary from $\sim 20$ Myr to $\sim 10$ Gyr.

Similarly, in the GC scenario, $\Psi_{\rm SGRB}(z)$ is given by convolution
of the GC formation rate, $\dot{\rho}_{\rm GC}(z)$, with a delay function
$f_{\rm GC}(t):$

\begin{equation}
\Psi_{\rm SGRB}(z) \propto \int_z^{\infty}dz' \frac{dt}{dz'}(z') \dot{\rho}_{\rm GC}(z') 
f_{\rm GC}[t(z)-t(z')].
\end{equation}

Hopman et al. (2006) considered $\dot{\rho}_{GC}$  proportional
to the observed star formation rate. Here we explore here three different
possibilities for $\dot{\rho}_{GC}$:
\begin{itemize}
\item model GC1: constant between $z=3$ and $z=5$ and null outside this interval,
as suggested from studies of GC formation in the 
cosmological context (Kravtsov \& Gnedin 2005);
\item model GC2: obtained from the GC age distribution of the spiral galaxy M31
(Fan et al. 2006);
\item model GC3: obtained from the GC age distribution of the elliptical galaxy 
M87 (Cohen, Blakeslee \& Ryzhov 1998).
\end{itemize}

For a comparison with Hopman et al. (2006) we explore a model
where $\dot{\rho}_{GC}$ follows the star formation rate, here computed
following Hopkins \& Beacom (2006), and referred as model GC4.

As regard to the function $f_{\rm GC}$, the delay time is the sum of
the interval between the formation of the GC and the dynamical
formation of the DNS, plus the DNS coalescence time due to gravitational 
wave emission.
The first time interval is set equal to the core-collapse time since,
only after the GC has experienced core-collapse, the rate of DNS
formation due to gravitational encounters becomes relevant (Hopman et
al. 2006). So the delay function $f_{\rm GC}(t)$ is the convolution of
the core--collapse time distribution and the 
gravitational wave coalescence time
distribution.  Following Hopman et al. (2006), the former is computed
from the compilation of the half-mass relaxation times and
concentrations of the observed GCs in our Galaxy by Harris (1996) \footnote{We
implicitly assume that the properties of GCs in Harris's catalogue are
representative of the population of GCs also in external galaxies.}
combined with the core-collapse time distribution from Quinlan (1996).

The latter is obtained from three-body simulations of encounters
between single neutron stars and target binaries 
in GCs. The resulting distribution
of eccentricities and orbital periods of the simulated DNS binaries
yields a delay function $\propto t^{-1.06}$ (Cerutti 2007) in close agreement with
Hopman et al. (2006). Characteristic delay times 
are longer than in the field case ($>3$ Gyr), and determined mainly by the 
extent of the core--collapse time. 

\section{The SGRB luminosity function}

For a comparison with data, we here compute the observed distribution of
SGRBs for the different DNS formation scenarios.

The observed photon flux, $P$, in the energy band 
$E_{\rm min}<E<E_{\rm max}$, emitted by an isotropically radiating source 
at redshift $z$ is

\begin{equation}
P=\frac{(1+z)\int^{(1+z)E_{\rm max}}_{(1+z)E_{\rm min}} S(E) dE}{4\pi d_L^2(z)},
\end{equation}

\noindent
where $S(E)$ is the differential rest--frame photon luminosity of the
source, and $d_L(z)$ is the luminosity distance.  To describe the
typical burst spectrum we adopt a single power--law with
an exponential cut--off with index  $-0.58$ 
and 
rest-frame break energy  $E_{\rm b}\sim 400$ keV (Ghirlanda, Ghisellini \& Celotti 2004)
where we assume a mean
redshift of $\sim 0.6$ for SGRBs.
We define the isotropic equivalent intrinsic 
burst luminosity as 
$L=\int^{2000\,\rm{keV}}_{30\,\rm{keV}} S(E)\, E\, dE$. 

Given a normalized SGRB luminosity function, $\phi(L)$, the observed rate of 
bursts with peak flux between $P_1$ and $P_2$ is

\begin{eqnarray}
\frac{dN}{dt}(P_1<P<P_2)&=&\int_0^{\infty} dz \frac{dV(z)}{dz}
\frac{\Delta \Omega_s}{4\pi} \frac{k_{\rm SGRB} \Psi_{\rm SGRB}(z)}{1+z}
\nonumber \\ & \times & \int^{L(P_2,z)}_{L(P_1,z)} dL^\prime
\phi(L^\prime),
\end{eqnarray}

\noindent
where $dV(z)/dz=4\pi c\, d_L^2(z)/[H(z)(1+z)^2]$ is the comoving volume
element\footnote{We adopted the concordance model values for the
cosmological parameters: $h=0.7$, $\Omega_M=0.3$, and
$\Omega_\Lambda=0.7$.}, and $H(z)=H_0 [\Omega_M
(1+z)^3+\Omega_\Lambda+(1-\Omega_M-\Omega_\Lambda)(1+z)^2]^{1/2}$.
$\Delta \Omega_s$ is the solid angle covered on the sky by the survey,
and the factor $(1+z)^{-1}$ accounts for cosmological time dilation.
$\Psi_{\rm SGRB}(z)$ is the comoving burst formation rate normalized to
unity at $z=0$ as computed in Section~2 and $k_{\rm SGRB}$ is the SGRB
formation rate at $z=0$.

In  this work, we assume a luminosity function described by a double power-law

\begin{equation}
\phi(L)\propto
\left\{
\begin{array}{rr}
(L/L_0)^{-\alpha} & \mbox{for } L<L_0 \\
(L/L_0)^{-\beta} & \mbox{for } L>L_0, \\
\end{array}
\right.
\end{equation}

\noindent
with $\alpha=1.6$ (Schmidt 2001) in order to reduce the number of
model free parameters.  The power index $\beta$ is a free parameter
of the model. The values obtained by fitting BATSE data are listed in
Table~1.

\begin{table}
\begin{center}
\begin{tabular}{lcccc}
\hline
\hline
Model & $k_{\rm SGRB}$ & $L_0/10^{50}$ & $\beta$ & $\chi^2_r$ \\
      & [Gpc$^{-3}$ yr$^{-1}$] & erg s$^{-1}$ & & \\
\hline
field & 12.7$\pm$4.6 & 1.16$\pm$0.01 & 1.41$\pm$0.22 & 1.1 \\
\hline
GC1 & 23.9$\pm$2.5 & 8.26$\pm$0.01 & 2.68$\pm$0.31 & 1.2 \\
GC2 &  90.2$\pm$11.6 & 3.22$\pm$0.01 & 2.72$\pm$0.37 & 1.2 \\
GC3 &  22.3$\pm$2.5 & 8.93$\pm$0.02 & 2.76$\pm$0.37 & 1.3 \\
GC4 & 37.6$\pm$2.3 & 6.81$\pm$0.01 & 2.80$\pm$0.29 & 1.3 \\
\hline
\hline
\end{tabular}
\end{center}
\caption{Best--fit parameters for the different DNS scenarios. 
Errors are at 1$\sigma$ level. The value of $k_{\rm SGRB}$ is obtained
in the hypothesis that the emission is isotropic (see Nakar 2007
for a discussion of the beaming factor).}
\end{table}

\begin{table*}
\begin{center}
\begin{tabular}{lccccl}
\hline
\hline
SGRB & $T_{90}$ & $P(64 {\rm ms})$ & $z$ & Used & References\\
    & [s] & [ph s$^{-1}$ cm$^{-2}$] & & & \\
\hline
050202 & $0.27$ & $4.0\pm0.4$ & & Y & 1\\
050509B & $0.073$ & $1.3\pm0.2$ & 0.226 & N & 2,3,4,5\\
050724 & $\sim 3^*$ &  $11.0\pm0.8$ & 0.257 & Y & 6,7,8,9,10,1,5\\
050813 & 0.45  & $2.1\pm0.21$ & 0.72 or 1.8 & Y & 11,1,5 \\
050906 & 0.258 & $1.7\pm0.3$ &  & N & 1\\
050925 & 0.07 & $10.8\pm1.0$ & & Y & 5\\
051105A & 0.093 & $2.7\pm0.5$ & & Y & 1\\
051210 & 1.3 & $1.0\pm0.2$ &  & N & 12,1,13,5\\
051221A & 1.4 & $40.7\pm1.5$ & 0.547 & Y & 14,1,13,5\\
060313 & 0.74 & $30.9\pm0.9$ & & Y & 15,1,16,13,5\\
060502B & 0.131 & $3.4\pm0.2$ & 0.287& Y & 17,1,13,5 \\
060801 & 0.49 & $2.1\pm0.2$ & 1.1304 & Y & 13\\
061006 & $\sim 0.42^*$ & $12.9\pm0.4$ & 0.4377 & Y & 13\\
061201 & 0.76 & $8.0\pm1.0$ & 0.111  & Y & 5\\
061210 & $\sim 0.19^*$ & $38.5\pm1.1$ & 0.41  & Y & 18 \\
061217 & 0.21 & $2.0\pm0.2$ & 0.827 & Y & 5\\
070209 & 0.09  & $2.8\pm0.4$ &   & Y & 19 \\
\hline
\end{tabular}
\end{center}
\caption{{\it Swift} SGRBs. $T_{90}$ from Sakamoto et al. (2007) apart from
the SGRBs denoted with $^*$ where we refer to the duration of the main pulse.
In the column `Used', we indicate with Y (N) the burst that we (don't) use  
in our analysis.
References (1), (3), (5), (9), (10), (12), (13), (15), (16), (19) are on the
short nature of the burst, whereas others refer to host and/or redshift
determination.
References: (1) Donaghy et al. (2006), (2) Gehrels et al. (2005), (3) Lee
et al. (2005), (4) Bloom et al. (2006b), 
(5) Berger (2007), (6) Barthelmy et al. (2005), 
(7) Berger et al. (2005), (8) Prochaska et al. (2006), (9) Campana et al.
(2006), (10) Grupe et al. (2006), (11) Berger (2005), (12) La Parola et al.
(2006), (13) Berger et al. (2007a), (14) Soderberg et al.
(2006), (15) Roming et al. (2006), (16) Postigo et al. (2006), (17) Bloom et al.
(2006a), (18) Berger (2006), (19) Sato et al. (2007)}
\end{table*}

\begin{figure}
\begin{center}
\centerline{\psfig{figure=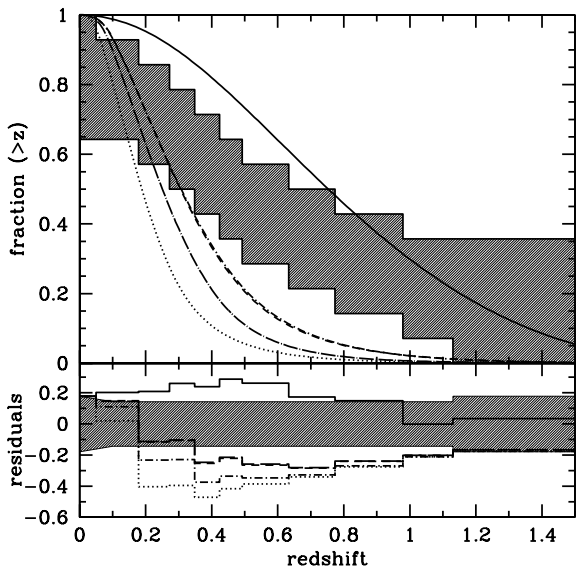,height=8cm}}
\caption{Shaded area: 
observed cumulative redshift distribution of {\it Swift} SGRBs
with $P_{64}>2$ ph s$^{-1}$ cm$^{-2},$ 
taking into account uncertainties due the lack of a secure redshift
determination for some SGRBs. Model results are overplotted for
different DNS formation channels: solid line refers to field scenario,
dot-short dashed to GC1, dotted to GC2, short dashed to GC3, and dot-long 
dashed to GC4.
In the bottom panel we plot the residuals to enhance the difference
between the model results and the observed distribution.
}
\label{low}
\end{center}
\end{figure}

We fit the differential photon flux distribution of SGRBs  with $T_{90}<2$ s 
detected by BATSE in the 50-300 keV band (Paciesas et al. 1999). We
consider the observed photon fluxes in the 64 ms timing window, 
restrict our fitting procedure to $P_{64}\ge 1$ ph s$^{-1}$ cm$^{-2}$
for which the detector response is efficient, and adopt a effective field of
view for BATSE, defined as $(\Delta \Omega_s/4\pi) t_{\rm obs},$  
of 1.8 yr. The best fit parameters are given in Table~1. 
We obtain a reasonable fit to BATSE data in all formation scenarios
by adjusting the free parameters of the model.
In this work 
we neglect the presence of 
a possible contamination from Soft Gamma-Ray Repeaters at low redshifts (Hurley
et al. 2005) which is still difficult to quantify (see Chapman et al. 2008). 

We then compute the expected SGRB rate in the 15--150 keV band of
{\it Swift} adopting a field of view  $\Delta \Omega_s=1.4$ sr, and a 2-year mission. 
For a correct comparison between model results and {\it Swift}
data, we calculate the photon peak flux for {\it Swift} SGRBs in a 
timing window similar to the one used on BATSE. 
Peak fluxes of Swift GRBs are calculated on
different integration times, spanning from 64 to 448 ms,
in order to match common significance criteria
and compatibly with the duration of the peak itself.
These are consistent within uncertainties with the
corresponding values integrated on a fixed 64-ms window.
This makes the comparison with BATSE well grounded.
The photon fluxes $P_{64}$ 
for the SGRBs observed with {\it Swift} are collected in Table~2.

The few {\it Swift} detections do not allow to construct a reliable
differential peak flux distribution. However,
we expect $\sim 9-10$ SGRBs
per year with $P_{64}\ge 2$ ph s$^{-1}$ cm$^{-2}$ for all our
models, in agreement with the 7 SGRBs per year detected by {\it Swift}
at the same flux limit. Note that this is a lower limit 
since we do not consider here 
SGRBs that are identified in the ground analysis (e.g., GRB~051114) 
and those long GRBs that could be  
tentatively ascribed to the short category (e.g., GRB~060614, and 060505). 
Lowering the {\it Swift} photon
flux threshold down to 1 ph s$^{-1}$ cm$^{-2}$, the expected SGRB detections largely
exceed the number of identified SGRBs. 
This fact may be related to a different energy band and trigger procedure 
of {\it Swift} and BATSE.
We thus choose the threshold of 2 ph 
s$^{-1}$ cm$^{-2}$ as measured in a 64 ms timing window as reliable detection 
limit for {\it Swift}.

\section{Redshift distribution}

In Table~2 we list all SGRBs detected by {\it Swift} until march  2007 with, when
present, the redshift. We note that GRB~050509B, 050906, and 051210
are not included in our analysis being their photon flux in the 64 ms
timing window below the threshold limit we adopted.  To be
conservative, we assume that GRB~050813 lies at $z=0.72$ instead of
$z=1.8$. The shaded area in Fig.~1 accounts for the observed
cumulative $z$--distribution including the uncertainties due the lack
of a secure redshift determination for some SGRBs.  Model results
computed for the different DNS formation channels are shown in
Fig.~1: solid line refers to the field scenario, dot--short dashed for GC1,
dotted for GC2, dashed for GC3, and dot--long dashed for GC4. For the field
scenario our fit is close to model (ii) of Guetta \& Piran (2007). 
For the GC scenario, we note that in spite of the
different assumptions of the GC formation rate, the expected SGRB
$z$--distributions are not remarkably different.  
The key result is that DNSs from dynamical interactions in
GC may account only for the lowest redshift SGRBs, i.e. $z<0.2-0.3$.
We find that only 4--12\% of SGRBs in GC should lie above $z=0.6$\footnote{
NS-NS binaries in young and dense star clusters
may form
dynamically and enhance
the rate of dynamical channel, at high redshifts
in star forming galaxies.}. This
result is expected owing to the longer delay times in the formation
and coalescence of DNSs in GCs relative to the field.  Field DNSs form
earlier and for this reason they can account for the SGRB distribution
above $z=0.8.$ Instead, the field scenario results inconsistent with
{\it Swift} data below this redshift. This appears to
be little affected by our choice of $\alpha$ in the luminosity function,
although the best--fit parameters in Table 1 may vary significantly.

Hopman et al. (2006) first noticed that SGRBs at low--$z$
may come from dynamical formation of DNSs in GCs 
and possibly ascribed all the SGRBs in the
{\it Swift} catalogue to this formation channels. We 
find here that {\it Swift} 2-year data
seem favour a bimodal origin of SGRBs:
the nearby, low--$z$ bursts resulting preferentially 
from dynamical forming DNSs, whereas distant, high-$z$
bursts resulting from the coalescence of primordial DNSs
following the cosmic star formation history.

Although current uncertainties are very large and the
data sample poor, we can try to quantify the relative weight of the two 
channels. In order to reproduce the observed limits on the redshift
distribution of SGRBs, almost 30\%--70\% of SGRBs detected by {\it Swift} 
should come 
from primordial NS-NS binary mergers if models GC1 or GC3 are considered.
For model GC2 this fraction increases up to 50--80\%. 
We can rewrite these results
in terms of the ratio between the intrinsic formation rates at $z=0$ 
in the two channels, i.e.
$\eta=k_{\rm SGRB,F}/k_{\rm SGRB,GC}$. We find that $\eta=0.22-1.24$ for
model GC1 and GC3, and $\eta=0.14-0.56$ for GC2. A large sample of SGRB
detection and host identifications are needed in order to confirm these
findings.

\section{Conclusions}

We have shown that current {\it Swift} data seem to point towards a
dual nature of SGRB formation: low--$z$ SGRBs may arise from the
coalescence of DNSs forming in GCs through dynamical collisions,
whereas high--$z$ bursts can represent the end--result of DNS mergers
born in the field from primordial massive binary stars. A way to
falsify this bimodality is by considering the correlation between SGRBs
and the host galaxy type (Gam-Yam et al. 2005; Zheng \& Ramirez-Ruiz 2007; 
Berger et al. 2007b; Shin \& Berger 2007).  We expect that SGRBs from the GC channel
should be found preferentially in early type galaxies where the bulk
of GCs resides, although we can not completely exclude a few in
spirals.  On the contrary, field SGRBs can be hosted in both early and
late type galaxies given the wide distribution of delay times (Section
2). Since the dynamical channel produces SGRBs at low--$z$, we expect
to find a excess of SGRBs in early type galaxies below $z<0.3$.  At
present it is still premature to exploit this issue in a statistically
meaningful manner and we may just confine ourselves to the analysis of
SGRBs in the current limited sample. Up to now, four SGRBs have been
localised in early-type galaxies (GRB~050509B, 050724, 050813, and
060502B), whereas six have secure late-type hosts (GRB~051221A,
060801, 061006, 061210, and 061217, Berger 2007; {\it HETE-II}
GRB~050709 at $z=0.161$; Fox et al. 2005; Covino et al. 2006).  
Fig.~2 shows the
fraction of SGRB hosts in early (shaded area) and late (white area)
below $z=0.3$ and above, respectively. We note indeed that low--$z$
SGRBs are associated preferentially to early-type galaxies while the
high--$z$ ones are located in late-type.  The large off--set of
GRB~050509B and 060502B may support further the association of these
SGRBs with DNSs formed dynamically in GCs.

\begin{figure}
\begin{center}
\centerline{\psfig{figure=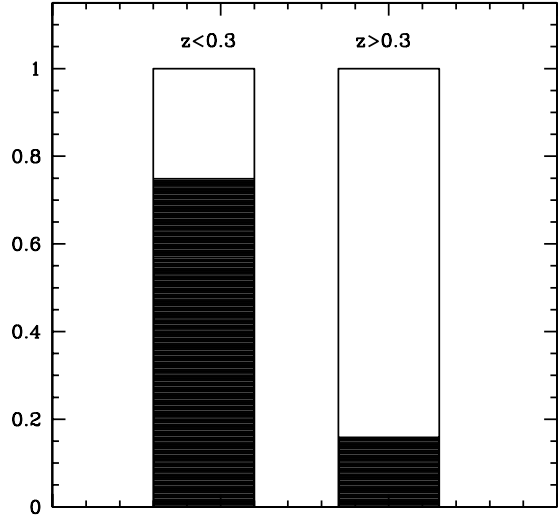,height=8cm}}
\caption{Fraction of SGRBs hosted in early--type 
(shaded area) and late--type (white area) galaxies  
below $z=0.3$ and above.}
\end{center}
\end{figure}

{}

\end{document}